\documentclass[12pt]{iopart}

\usepackage{graphicx}
\usepackage{color}
\usepackage{dcolumn}
\usepackage{bm}
\usepackage{amssymb}

\begin{document}


\title[Coherent and incoherent processes in magneto-optical signals]{Coherent and incoherent processes responsible for various characteristics of nonlinear magneto-optical signals in rubidium atoms}

\author{Marcis Auzinsh, Andris Berzins, Ruvin Ferber, Florian Gahbauer, Linards Kalvans, Arturs Mozers}
\ead{\mailto{mauzins@latnet.lv}}%
\address{%
The University of Latvia, Laser Centre, Rainis Boulevard 19, LV-1586 Riga, Latvia
}%

\date{\today}

\begin{abstract}
 We present the results of an investigation of the different physical processes that influence the shape of the nonlinear magneto-optical signals both at small magnetic field values ($\sim\! 100$~mG) and at large magnetic field values (several tens of Gauss). We used a theoretical model that provided an accurate description of experimental signals for a wide range of experimental parameters. By turning various effects ``on'' or ``off'' inside this model, we investigated the origin of different features of the measured signals. We confirmed that the narrowest structures, with widths on the order of 100 mG, are related mostly to coherences among ground-state magnetic sublevels. The shape of the curves at other scales could be explained by taking into account the different velocity groups of atoms that come into and out of resonance with the exciting laser field. Coherent effects in the excited state can also play a role, although they mostly affect the polarization components of the fluorescence. The results of theoretical calculations are compared with experimental measurements of laser induced fluorescence from the $D_2$ line of atomic rubidium as a function of magnetic field.  
\end{abstract}

\maketitle

\section{\label{sec:level1}Introduction}
	When coherent radiation excites an atomic system with ground-state angular momentum $F_g$ and excited-state angular momentum $F_e$, coherences can be created among the magnetic sublevels~\cite{Aleksandrov:1993, optically-polarized-atoms}. At low laser intensity, coherences appear in the excited state of the atom. As the laser intensity increases, the absorption processes become nonlinear, and coherences are created among the magnetic sublevels of the ground state as well. When the degeneracy among the magnetic sublevels is lifted by applying an external field (in our case magnetic), the coherences are destroyed. As a result, nonlinear magneto-optical resonances (NMOR) can be observed in the laser-induced fluorescence (LIF) plotted as a function of magnetic field. 
	For linearly polarized radiation exciting a transition $F_g\longrightarrow F_e=F_g+1$, these resonances will be bright, that is, the atoms will be more absorbing at zero magnetic field~\cite{Dancheva:2000,Kazantsev:1984,Renzoni:2001b,AlnisJPB:2001}. When $F_e \leq F_g$, the resonances will be dark, or less absorbing at zero magnetic field~\cite{Schmieder:1970,Alzetta:1976}.
	The NMOR features can be as narrow as $10^{-6} - 10^{-5}$~G when buffer gas or antirelaxation coating of the cell is used because of the slow relaxation rate of the ground state~\cite{Budker:1998}. This characteristic makes them suitable for many applications, such as, for example, magnetometry~\cite{Scully:1992}, lasing without inversion~\cite{Scully:1989}, electrically induced transparency~\cite{Harris:1997}, slow light and optical information storage~\cite{Phillips:2001, Liu:2001}, atomic clocks~\cite{Knappe:2005}, and narrow-band optical filters~\cite{Cere:09}.
	However, these narrow resonances are usually found within broader structures with features on the order of several Gauss or several tens of Gauss in a plot of LIF versus magnetic field. Our study focuses on these broader structures, which are interesting in themselves and also for some practical applications at higher magnetic field values, like optical isolators~\cite{Weller:12}. Using a theoretical model that has been developed over time and mostly was used to describe the narrow magneto-optical resonances but can reproduce the magneto-optical signals with high accuracy over a large range of magnetic field values~\cite{Auzinsh:13}, we investigated the peculiar shape and sign (bright or dark) of these structures, as well as the physical processes that give rise to them.  
	
 In order to describe magneto-optical signals over a magnetic field range of several tens of Gauss or more, it is necessary to include in the model excited-state coherences, energy shifts of the magnetic sublevels in external fields, which bring levels out of resonance with the narrow-linewidth laser radiation, and the magnetic-field-induced mixing of the atomic wavefunctions, which changes the transition probabilities of the different transitions between ground and excited-state sublevels~\cite{Sargsyan:2008, Sargsyan:2012}. Moreover, it is necessary to treat various relaxation processes, the coherence properties of the laser radiation, and the Doppler effect. Since at least the 1970s, magneto-optical signals in alkali atoms have been modelled by solving the optical Bloch equations for the density matrix~\cite{Picque:1978}. Simple models were able to describe the narrow resonances fairly well~\cite{Papoyan:2002}, but failed to describe the signals at fields of several Gauss or more. With time, these models become more sophisticated as the aforementioned effects were incorporated ~\cite{Andreeva:2002,Blushs:2004,Auzinsh:2008}, and now the agreement is often excellent, at least up to magnetic fields over one hundred Gauss. Thus, numerical models have become useful tools for understanding the physical processes that give rise to various features in the signals, because different physical processes can be included in the models or excluded one by one. Analytical studies, on the other hand, can demonstrate more explicitly a link between a particular physical process and the observable outcome. Thus, in \cite{Papoyan:2002} analytical formulae were developed that allow one to calculate the contrast of bright resonances. In another study, a theoretical model of the electromagnetically induced absorption (EIA) was constructed for a hypothetical $F_g=1\longrightarrow F_e=2$ transition~\cite{Brazhnikov:2005}. It was possible to show from a purely theoretical point of view that the sub-natural linewidth resonance in EIA was related to the transfer of coherence from the excited state to the ground state. More recently, sophisticated analytical models were developed that are 
valid in the low-power region, and were applied to experimental measurements on the caesium $D_1$ line~\cite{Castagna:2011,Breschi:2012}. Comparison with experiments confirmed that the narrow resonances arise when polarization is transferred from the excited state to the ground state. In~\cite{Auzinsh:2009-2} an analytical model was used to analyze the influence of partially resolved hyperfine structure in the ground or excited state on nonlinear magneto-optical rotation signals.
Numerical studies such as ours can complement these analytical investigations, because the numerical models can be made to apply over a wider range of laser power densities and consider realistic, Doppler-broadened atomic transitions in the manifold of the hyperfine levels, that is to say, take into account multiple adjacent transitions.    

	Our study focused on the $D_2$ line of $^{87}$Rb as a model system. Since the origin of the narrow structure had already been shown to be connected to coherences in the ground state~\cite{Brazhnikov:2005,Breschi:2012}, our study primarily aimed at understanding the wider features of the magneto-optical signals up to magnetic field values of several tens of Gauss, since such understanding is important in itself and will help to improve the models of the narrow resonances used for applications.  Nevertheless, since a numerical model such as ours gives complete flexibility to turn different effects ``on'' and ``off'', we were also able to confirm the origin of the narrow structure using a different technique, \textit{i. e.}, one that is not analytical.
	
\begin{figure}
\includegraphics{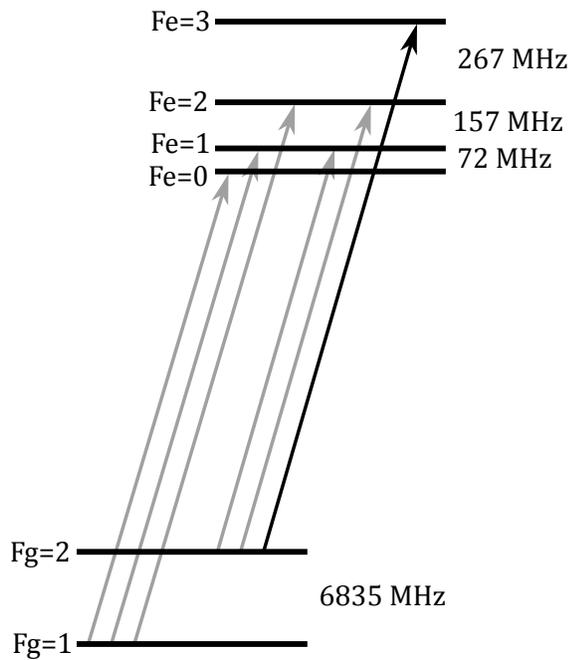}
\caption{\label{fig:level-scheme} Scheme of the hyperfine levels and allowed transitions of the $D_2$ line of $^{87}$Rb. }
\end{figure}

\begin{figure}
\includegraphics{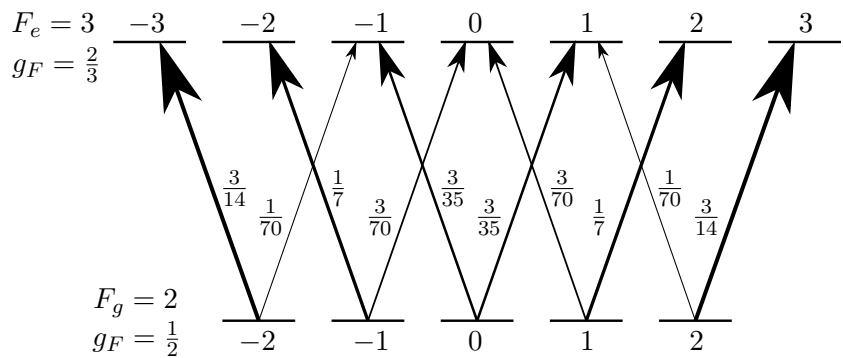}
\caption{\label{fig:probabilities}  Relative transition strengths from the ground-state magnetic sublevels to the excited-state magnetic sublevels when the linearly polarized exciting radiation is decomposed into $\sigma^\pm$ circularly polarized components for the $F_g=2\longrightarrow F_e=3$ transition of the $D_2$ line. The Lande factor $g_F$ is given at the left of each particular hyperfine level}
\end{figure}

\begin{figure}
\resizebox{2in}{!}{\includegraphics{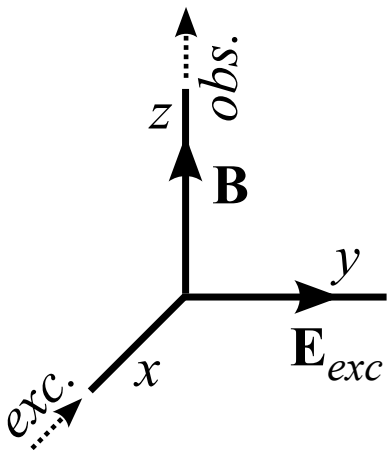}}%
\caption{\label{fig:geometry} Geometry of the excitation and observation directions.}
\end{figure}

	The level structure of the transition studied here is shown in Fig.~\ref{fig:level-scheme}~\cite{Steck:rubidium87}. The transition was excited by  linearly polarized laser radiation. Figure~\ref{fig:probabilities} shows the relative transition probabilities from the ground-state sublevels of the $F_g=2$ level to the excited-state sublevels of the $F_e=3$ level when the linearly polarized exciting radiation is decomposed into coherent circularly 
polarized 
components. It is assumed that the light is polarized perpendicularly to the direction of the external magnetic field (see Fig.~\ref{fig:geometry}.) This scheme implies that $\Delta m = 2$ coherences are created between different Zeeman sublevels in the excited state as well as in the ground state. Two distinct processes contribute to ground-state coherence. The first process creates coherence in the ground state through direct interaction with the radiation field via $\Lambda$-type absorption. 
In the second proccess the $V$-type absorption creates coherences in the excited state, which then can 
be transferred back to the ground state via spontaneous emission, see Eq. (13.13) in~\cite{optically-polarized-atoms}.
Fig.~\ref{fig:probabilities} shows that both V-type and $\Lambda$-type transitions are present in our physical system. 

The paper is organized as follows: Sec. II outlines the theoretical model. In Sec. III we describe the experimental conditions, and in Sec. IV we discuss the results and attempt to decompose the modelled signal into components that are related to different physical processes. 

\section{Theoretical Model}
\label{sec:theory}
The theoretical model is based on the density matrix approach. The density matrices are written in the $\vert \xi, F_i, m_F\rangle$ basis where $F_i$ denotes the quantum number of the total atomic angular momentum, $m_F$, the respective magnetic quantum number, and $\xi$, all other quantum numbers. The time evolution of the density matrix is described by the optical Bloch equations~\cite{Stenholm:2005}
\begin{equation}\label{eq:liouville}
	i\hbar\frac{\partial \rho}{\partial t} = \left[\hat H,\rho\right] + i\hbar\hat R\rho,
\end{equation}
which include the full atomic Hamiltonian $\hat H = \hat H_0 + \hat H_B + \hat V$ constructed from the unperturbed atom's Hamiltonian $\hat H_0$, which depends on the internal dynamics of the atom, the Hamiltonian $\hat H_B$, which describes the atom's interaction with the external magnetic field, and the dipole operator $\hat V$, which represents the atom's interaction with the electromagnetic radiation. The interaction with the magnetic field gradually decouples the total electronic angular momentum $\mathbf{J}$ and nuclear spin $\mathbf{I}$, which means that $F$ no longer is a good quantum number, while $m$ still remains a good quantum number. To deal with this effect, mixing coefficients between different hyperfine states in the magnetic field are introduced in the model. The relaxation operator $\hat R$ in (\ref{eq:liouville}) accounts for the spontaneous decay that transfers atoms from the excited state to the ground state, the collisional relaxation, and the transit relaxation. The latter occurs when atoms leave and enter the interaction region as a result of their thermal motion.

The optical Bloch equations can be written explicitly for each element of the density matrix. Applying the rotating wave approximation and assuming the density matrices do not follow promptly the random phase fluctuations of the electromagnetic radiation, we may decorrelate the time-dependent differential equations from the fluctuating phase and average over it. Thus we may adiabatically eliminate the equations that describe the optical coherences and obtain rate equations for the Zeeman coherences~\cite{Blushs:2004}:
\numparts\label{eq:zc}
\begin{eqnarray}
\fl\frac{\partial \rho_{g_ig_j}}{\partial t} = \left(\Xi_{g_ie_m} + \Xi_{g_je_k}^{\ast}\right)\sum_{e_k,e_m}d_{g_ie_k}^\ast d_{e_mg_j}\rho_{e_ke_m} - \nonumber\\ - \sum_{e_k,g_m}\Big(\Xi_{g_je_k}^{\ast}d_{g_ie_k}^\ast d_{e_kg_m}\rho_{g_mg_j} + \Xi_{g_ie_k}d_{g_me_k}^\ast d_{e_kg_j}\rho_{g_ig_m}\Big) - \nonumber\\- i \omega_{g_ig_j}\rho_{g_ig_j} - \gamma\rho_{g_ig_j} + \sum_{e_ke_l}\Gamma_{g_ig_j}^{e_ke_l}\rho_{e_ke_l} + \lambda\delta(g_i,g_j) \label{eq:zcgg} \\
\fl\frac{\partial \rho_{e_ie_j}}{\partial t} = \left(\Xi_{g_me_i}^\ast + \Xi_{g_ke_j}\right)\sum_{g_k,g_m}d_{e_ig_k} d_{g_me_j}^\ast\rho_{g_kg_m} -\nonumber\\- \sum_{g_k,e_m}\Big(\Xi_{g_ke_j}d_{e_ig_k} d_{g_ke_m}^\ast\rho_{e_me_j} + \Xi_{g_ke_i}^\ast d_{e_mg_k} d_{g_ke_j}^\ast\rho_{e_ie_m}\Big) - \nonumber\\ - i \omega_{e_ie_j}\rho_{e_ie_j} - (\Gamma + \gamma)\rho_{e_ie_j}. \label{eq:zcee}
\end{eqnarray}
\endnumparts
In both equations of (\ref{eq:zc}) the first term describes the optically induced transitions to the level described by a particular equation, and the second term, the transitions away from it, with $d_{ij}$ being the element of the dipole transition matrix that can be calculated according to the Wigner-Eckart theorem~\cite{optically-polarized-atoms}. The terms $\Xi_{g_ie_j}$ and complex conjugate $\Xi^\ast_{e_jg_i}$ are described below. The third term describes the coherence destruction by the magnetic field, with $\omega_{ij} = \frac{E_i-E_j}{\hbar}$ denoting the energy difference between levels $\vert i\rangle$ and $\vert j\rangle$ caused by both the hyperfine splitting and the nonlinear Zeeman effect. The fourth term describes relaxation due to transit relaxation, collisions  and spontaneous decay (only for the excited state). Two additional terms in (\ref{eq:zcgg}) stand for population transfer to the ground state via spontaneous decay from the excited state (fifth term) and unpolarized atoms entering the interaction region as a result of their thermal motion (sixth term).

The symbol $\Xi_{g_ie_j}$ in equation (\ref{eq:zc}) describes the strength of interaction between the laser radiation and the atoms and is expressed as follows:
\begin{equation}\label{eq:xi}
\Xi_{g_ie_j} = \frac{\Omega_{R}^{2}}{\frac{\Gamma+\gamma+\Delta\omega}{2}+\dot\imath\left(\bar\omega-\mathbf{k}_{\bar\omega}\mathbf{v}
	+\omega_{g_ie_j}\right)},
\end{equation}
where $\Omega_{R}$ is the Rabi frequency, further discussed in Sec.~\ref{sec:results}, $\Gamma$ and $\gamma$ are the rates of spontaneous decay and transit relaxation, $\Delta\omega$ is the finite spectral width of the exciting radiation, $\bar \omega$ is the central frequency of the exciting radiation, $\mathbf{k}_{\bar\omega}$ the respective wave vector, and $\mathbf{k}_{\bar\omega}\mathbf{v}$ is the Doppler shift experienced by an atom moving with a velocity $\mathbf{v}$. The dependence of the absolute value of $\Xi_{g_ie_j}$ at fixed $i$ and $j$ on the magnetic field is responsible for the effects of magnetic scanning discussed in Sec. \ref{sec:results}, while the imaginary part of $\Xi_{g_ie_j}$ represents the dynamic Stark effect.

The steady state solution of the rate equations (\ref{eq:zc}) yields the density matrices that describe population of magnetic sublevels and Zeeman coherences of both the ground and excited states. The density matrix of the excited state is used to calculate the fluorescence signal for an arbitrary polarization component $\mathbf{e}$:
\begin{equation}
	I_{fl}(\mathbf{e}) = \tilde{I}_0\sum\limits_{g_i,e_j,e_k} d_{g_ie_j}^{\ast(ob)}d_{e_kg_i}^{(ob)}\rho_{e_je_k},
\end{equation}
where $\tilde{I}_0$ is a proportionality coefficient and $d_{e_jg_i}^{(ob)}$ are elements of the dipole transition matrix for the chosen observation component $\mathbf{e}$. The unpolarized fluorescence signal in a particular direction was calculated by summing over two orthogonal polarization components. To take into account the Doppler effect, this quantity
was averaged over the one-dimensional Maxwellian distribution of atomic velocity along the direction of the laser beam propagation axis.  In addition, the density matrices for some particular velocity groups are used to obtain angular momentum probability surfaces~\cite{optically-polarized-atoms, Auzinsh:1997,Rochester:2001}.

\section{Experiment}
\label{sec:experiment}
	The experiments were carried out at room temperature on natural mixture of rubidium isotopes in a cylindrical Pyrex vapour cell with optical quality windows, $25$~mm long and $25$~mm in diameter, produced by Toptica, A.G. of Graefelfing, Germany.  The geometry of the excitation and observation is shown in Fig.~\ref{fig:geometry}. The $780$~nm exciting laser radiation propagates along the $x$ axis with linear polarization vector $\mathbf{E}$ pointing along the $y$ axis. The total LIF (without polarization or frequency discrimination) was observed along the $z$ axis, which was parallel to the magnetic field vector $\mathbf{B}$. The laser was a home-made extended-cavity diode laser. The magnetic field was supplied by a Helmholtz coil and its value was scanned by controlling the current in a Kepco BOP-50-8-M bipolar power supply. Signals were recorded by a photodiode (Thorlabs FDS-100). The laser frequency was determined by means of a saturated spectroscopy setup in conjunction with a wavemeter (WS-7 made by HighFinesse). The beam profile was measured by means of a beam profiler (Thorlabs BP104-VIS). The full width at half maximum was assumed to be the beam diameter used in the calculations [see Eq.~(\ref{eq:gamma}), Sec.~\ref{sec:results}].
The ambient magnetic field along the $x$ and $y$ directions was compensated by a pair of Helmholtz coils. The entire experimental setup was located on a nonmagnetic optical table. Possible inhomogeneity of the magnetic field along the laser propagation axis might be caused by imperfect Helmholtz coils and does not exceed $13$ $\mu$G according to an estimation based on coils' dimensions.

\section{Results and Discussion}
\label{sec:results}
\begin{figure*}[ht] 
\resizebox{.5\textwidth}{!}{\includegraphics[width=.5\textwidth]{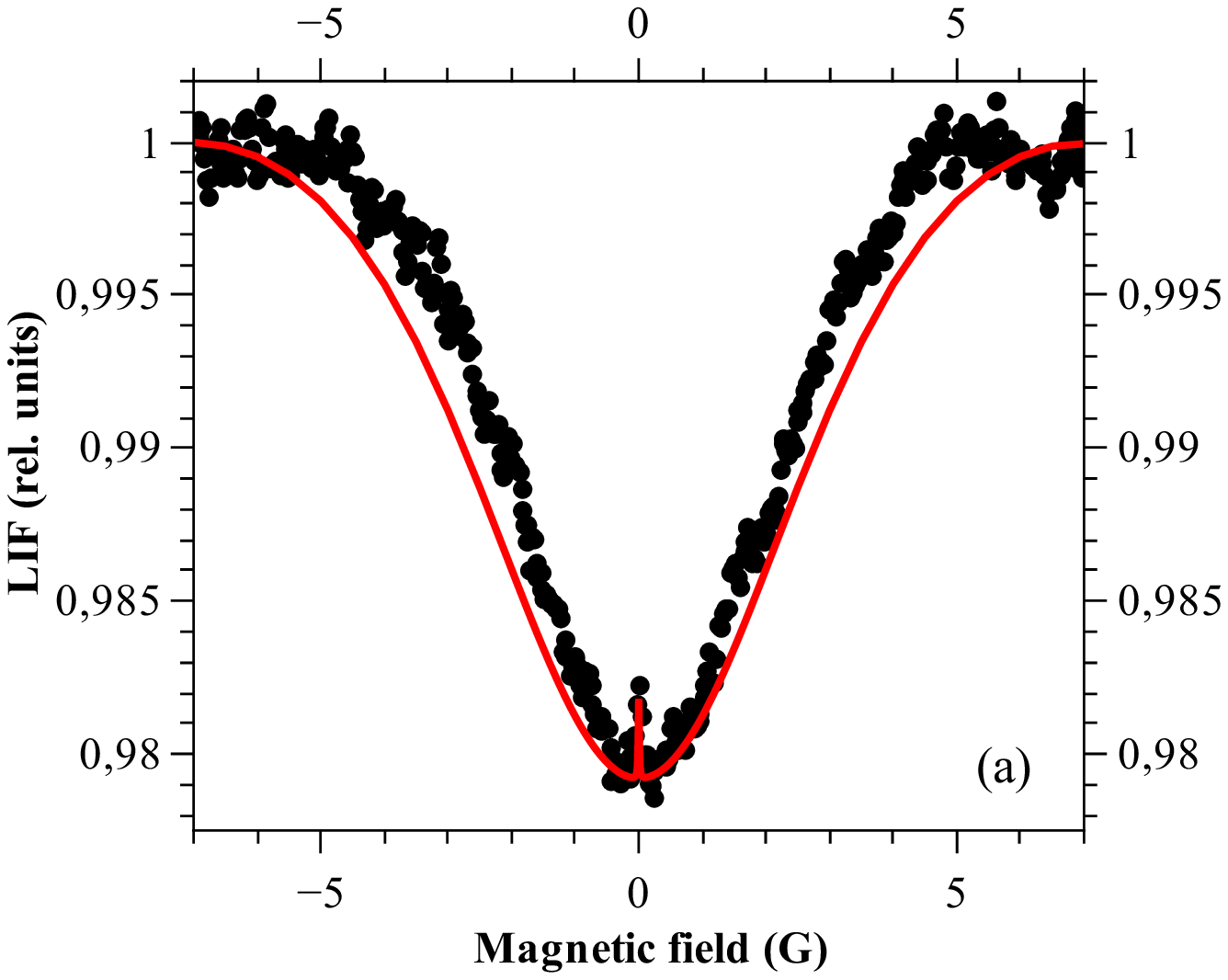}}%
\resizebox{.5\textwidth}{!}{\includegraphics[width=.5\textwidth]{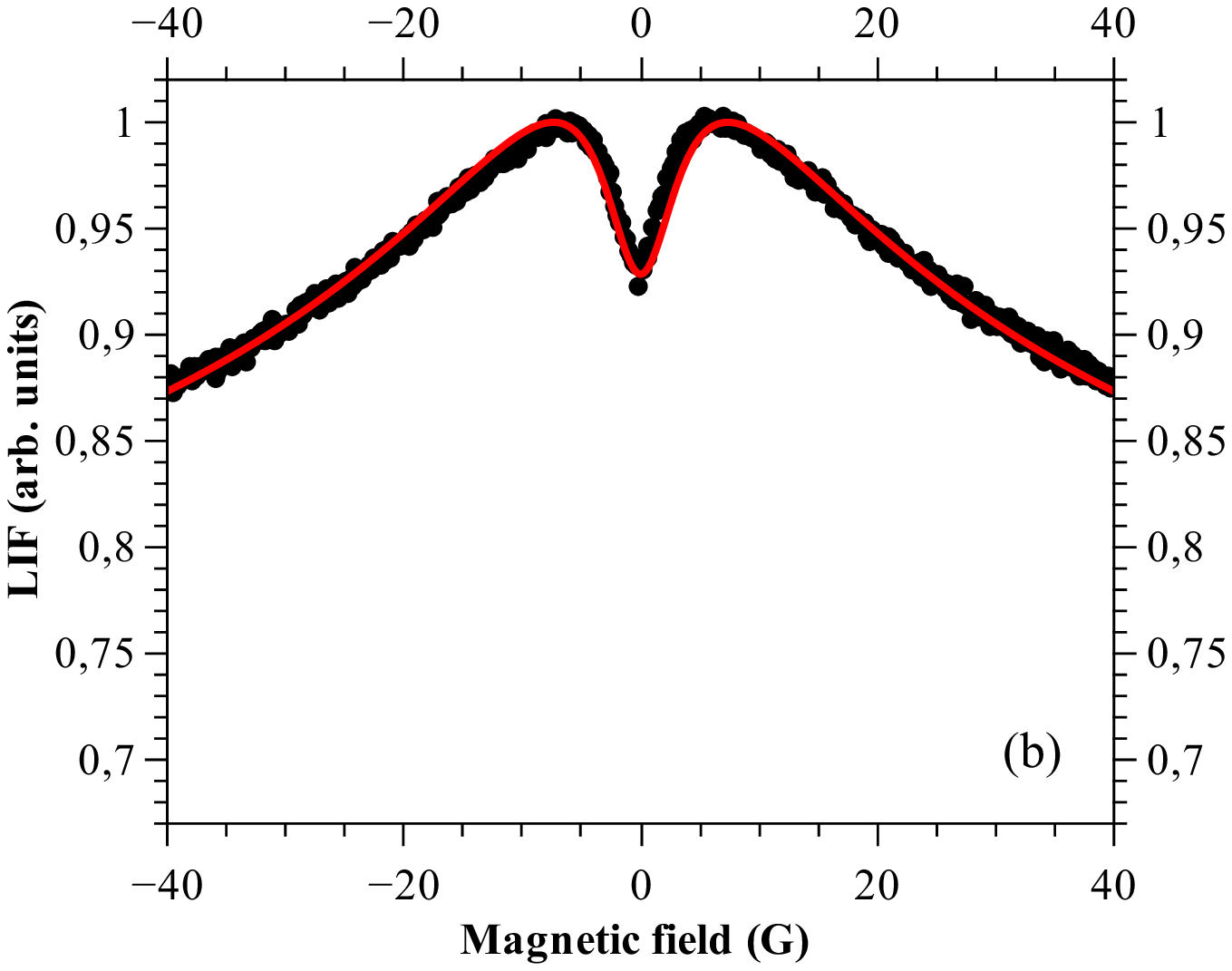}}\\%
\resizebox{.5\textwidth}{!}{\includegraphics[width=.5\textwidth]{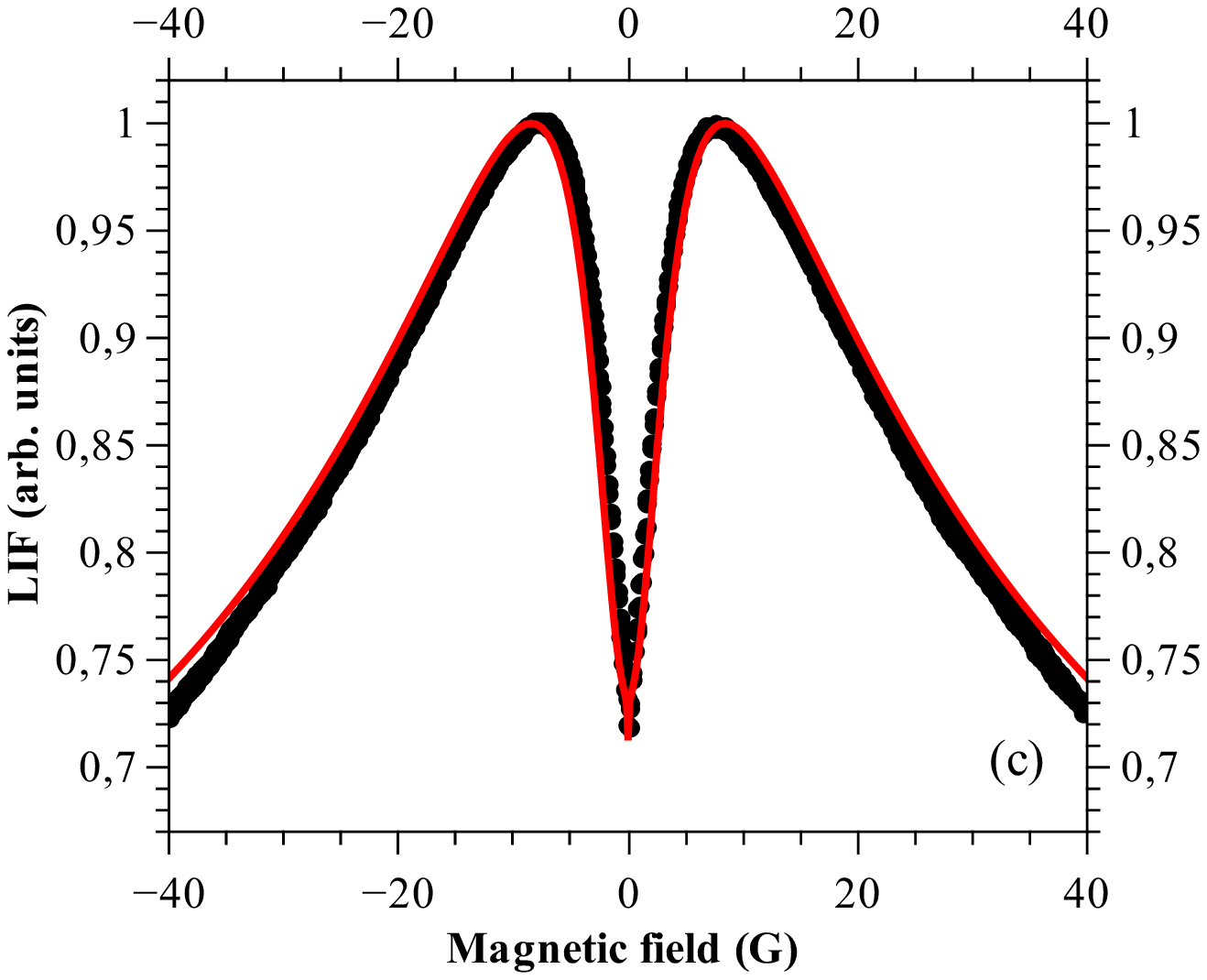}}%
\resizebox{.5\textwidth}{!}{\includegraphics[width=.5\textwidth]{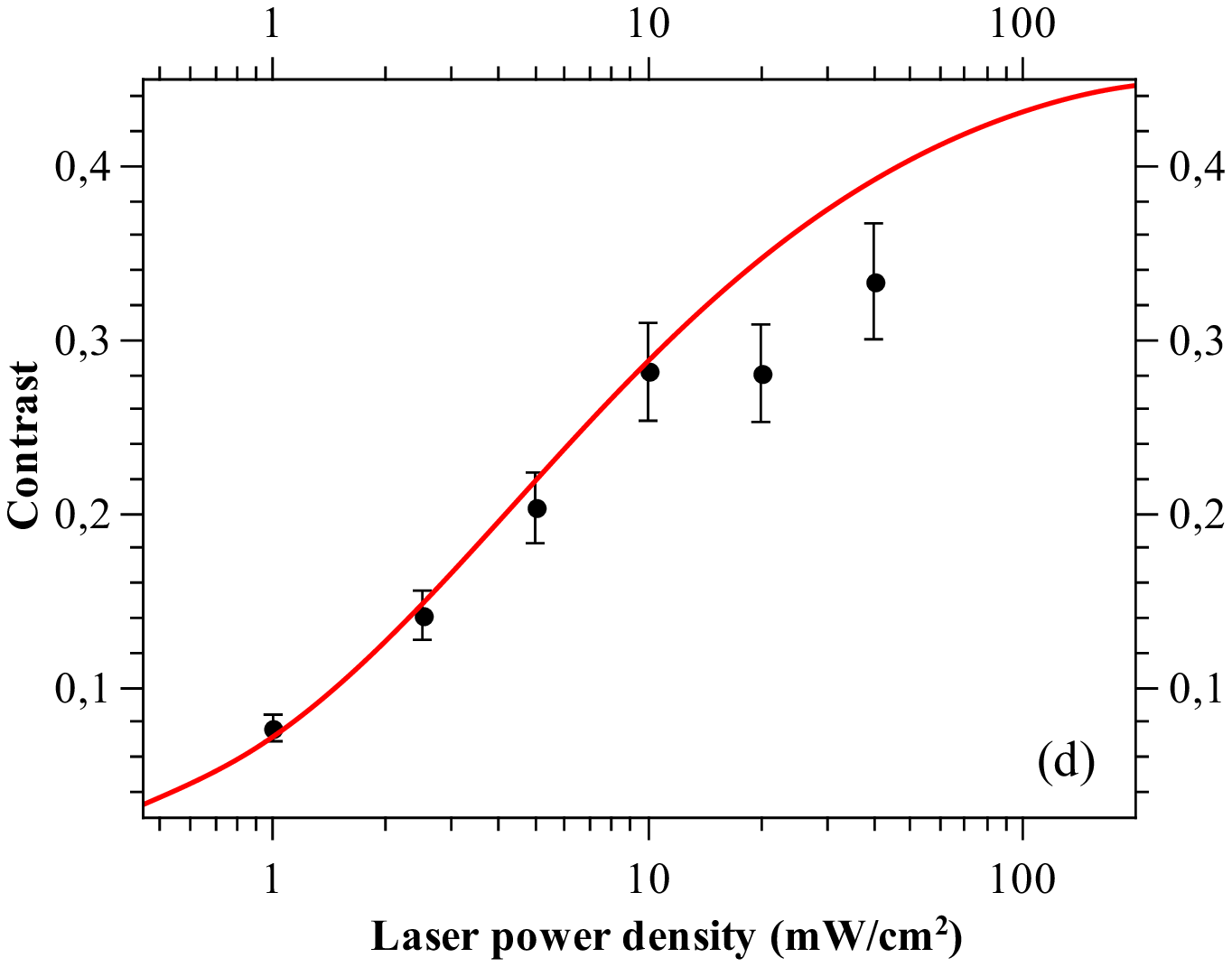}}%
\caption{\label{fig:profiles} (Colour online) LIF versus magnetic field value for the $F_g=2\longrightarrow F_e=3$ transition of $^{87}$Rb for different values of the laser power density $I$: (a) 0.14 mW/cm$^2$, (b) 1 mW/cm$^2$, (c) 10 mW/cm$^2$. The bottom right panel shows the contrast of the central minimum as a function of laser power density. Filled circles correspond to experimentally measured values, whereas the solid line shows the result of a calculation. Note the different scales in (a) and (b-c).}
\end{figure*}

	As the main tool for our present investigation was a numerical model, the first step was to show that it accurately described the measured signals over a large range of magnetic field values. Previous studies had already shown the model to be accurate in many experimental situations~\cite{Auzinsh:2012} in which narrow magneto-optical resonances form in weak magnetic fields ($B \lesssim 0.3$~G) as a result of coherences created among the magnetic sublevels of the ground state. Figure~\ref{fig:profiles} shows plots of LIF versus magnetic field over the range $-40$~G to $+40$~G when the laser was tuned to the $F_g=2\longrightarrow F_e=3$ transition of the $D_2$ line of $^{87}$Rb at different laser power densities, as well as a plot of contrast versus laser power density. It must be noted that due to proximity of other hyperfine levels in the excited state, the Doppler effect and magnetic scanning, other hyperfine levels were also excited at least partially. These transitions are included in our theoretical model as well. We defined the signal contrast as
\begin{equation}
	C = \frac{I_{min} - I_{max}}{I_{max}},
\end{equation}
where $I_{min}$ is the minimum LIF value (zero first derivative and positive second derivative) around $B=0$, and $I_{max}$ is the LIF value at the first point with vanishing first derivative and $\vert B\vert > 1$~G.
	Filled circles represent experimentally measured values, whereas the line shows the result of a theoretical calculation.  In order to obtain an appropriate fit to the data, it was necessary to adjust two parameters. The first 
parameter was the constant $k_{\gamma}$ that relates the ratio of the mean thermal velocity $v_{th}$ of the atoms and the characteristic diameter of the laser beam $d$ to the transit relaxation rate $\gamma$ as
\begin{equation}
	\gamma = k_{\gamma} \frac{v_{th}}{d} + \gamma_{col} + \gamma_{hom} \approx k_{\gamma} \frac{v_{th}}{d},
	\label{eq:gamma}
\end{equation}
where $\gamma_{col}$ is the rate of inelastic atom--atom collisionsand $\gamma_{hom}$ is the relaxation caused by inhomogenities of the magnetic field. An estimated value of $\gamma_{col}$ at room temperature, assuming the spin-exchange cross section for Rb-Rb collisions ${\sigma\approx 2\cdot 10^{-14}}$~cm$^2$~\cite{Happer:1972} is several orders of magnitude less than the first term in (\ref{eq:gamma}). The upper limit of $\gamma_{hom}$ estimated as shown in~\cite{Pustelny:2006} is also several orders of magnitude less than the first term. So both, $\gamma_{col}$ and $\gamma_{hom}$ were omitted in the actual calculations.
The second parameter $k_R$ related the Rabi frequency $\Omega_R$ to the square root of the the experimental laser power density $I$ according to
\begin{equation}
	\Omega_R = k_{R}\frac{\vert\vert d\vert\vert}{\hbar}\sqrt{\frac{2I}{c}},
\end{equation}
where $\vert\vert d\vert\vert$ is the reduced dipole matrix element that remains unchanged for all transitions within the $D_2$ line at a well documented value~\cite{optically-polarized-atoms}, and $c$ is the speed of light.
Both fitting parameters ($k_{\gamma}$ and $k_{R}$) would be equal to unity for a rectangular beam profile of the exciting laser and atoms moving with the mean thermal velocity across the middle of the beam profile. In our experiment the beam profile is roughly Gaussian, and so the laser beam diameter cannot be defined unambiguously. Furthermore, atoms are moving along random trajectories with velocities distributed according to the Maxwellian velocity distribution. Thus we allow the values of these constants to deviate from unity in order to obtain an optimal fit between the modelled and experimentally recorded results. A full numerical integration over both (Gaussian and Maxwellian) distributions would be too time consuming, while our approach has proven to describe experimental results with high accuracy in previous studies e.g.~\cite{Auzinsh:2012, Auzinsh:13}.

The actual values of the fitting parameters were $k_\gamma = 0.5$ and $k_R = 0.11$. These values indicate that the interaction of atoms and laser radiation in the wings of the (roughly Gaussian) beam profile cannot be neglected, please see~\cite{Auzinsh:2012-2} for more detailed discussion.
Thus for a beam with $d = 1.6$~mm (estimated in the experiment as defined in Sec.~\ref{sec:experiment}) and laser power $P = 20$~$\mu$W, we obtained the following values that were used in the modelling: $\gamma = 95$~kHz and $\Omega_R = 0.75$~MHz. Another important parameter for modelling and interpreting the results is the natural linewidth, which is $\Gamma = 6.067$~MHz~\cite{Steck:rubidium87}.
Having obtained the optimum values for these parameters by trial and error, these values were used to fit simultaneously all experimental data obtained for different transitions and different values of the laser power density. (The top left plot in Fig. \ref{fig:profiles} was measured in a different experiment dedicated to the narrow structure~\cite{Auzinsh:2012}, and so the experimental conditions and fitting parameters were slightly different in this case, and the range of the measured magnetic field was smaller.) Agreement between experiment and theory was rather satisfactory, which shows that the model serves as a good basis for understanding the dependence of LIF on the magnetic field over a broad range of magnetic field values.

The narrow resonance at zero magnetic field is related to the destruction of coherences in the ground state by the magnetic field as we will show in the next paragraphs. Under our experimental conditions, it had a width of about one hundred milligauss and was clearly visible right at zero magnetic field. A detailed study of this resonance was performed in~\cite{Auzinsh:2012} showing that this structure points up or down (changes the sign of the second derivative) depending of the laser power density. Under the present experimental conditions it appeared as a narrow structure with a negative second derivative (pointing upwards). This narrow resonance was located in the center of another structure with a positive second derivative and a width of several Gauss. 

\begin{figure*}[ht] 
\resizebox{.5\textwidth}{!}{\includegraphics[width=.5\textwidth]{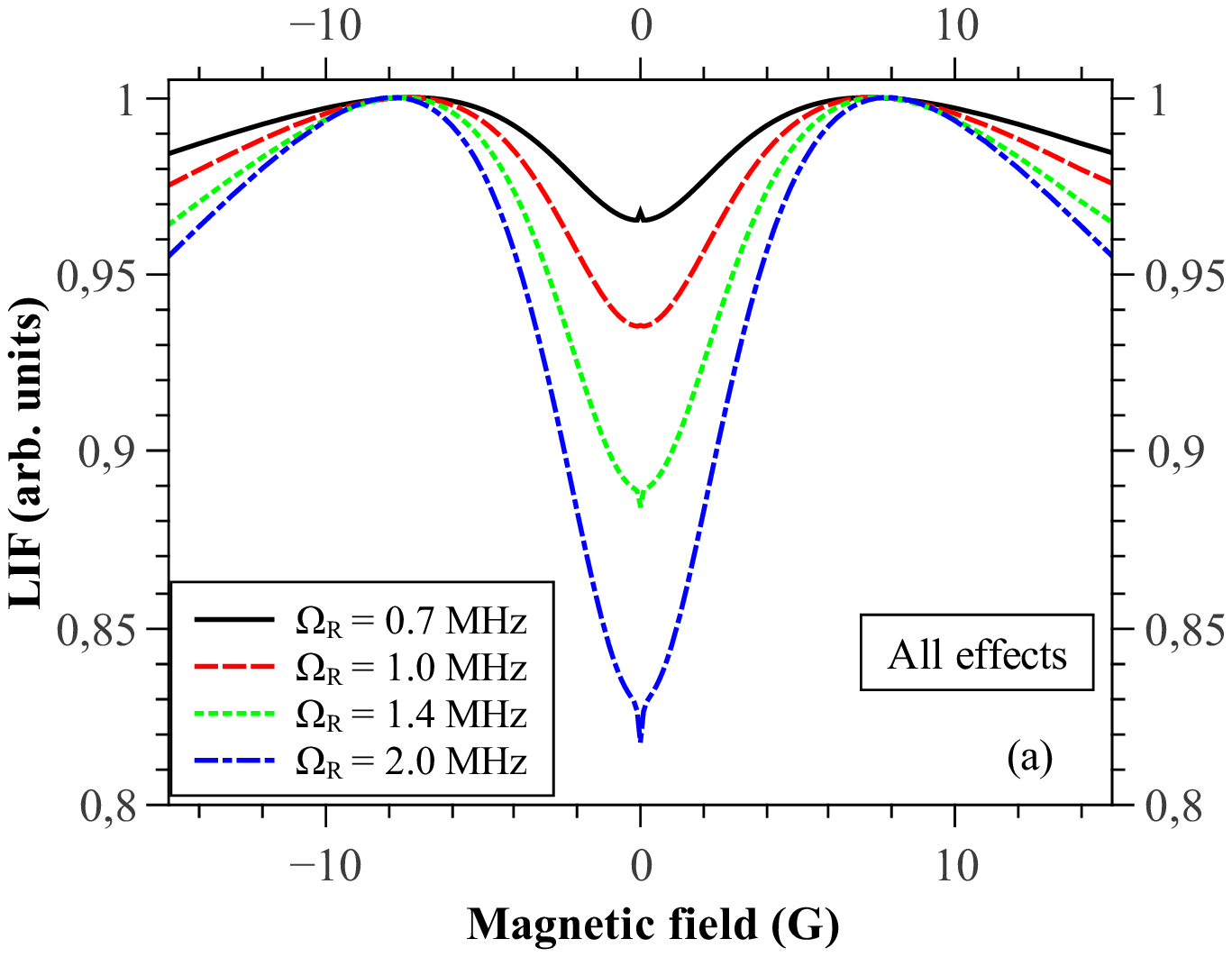}}%
\resizebox{.5\textwidth}{!}{\includegraphics[width=.5\textwidth]{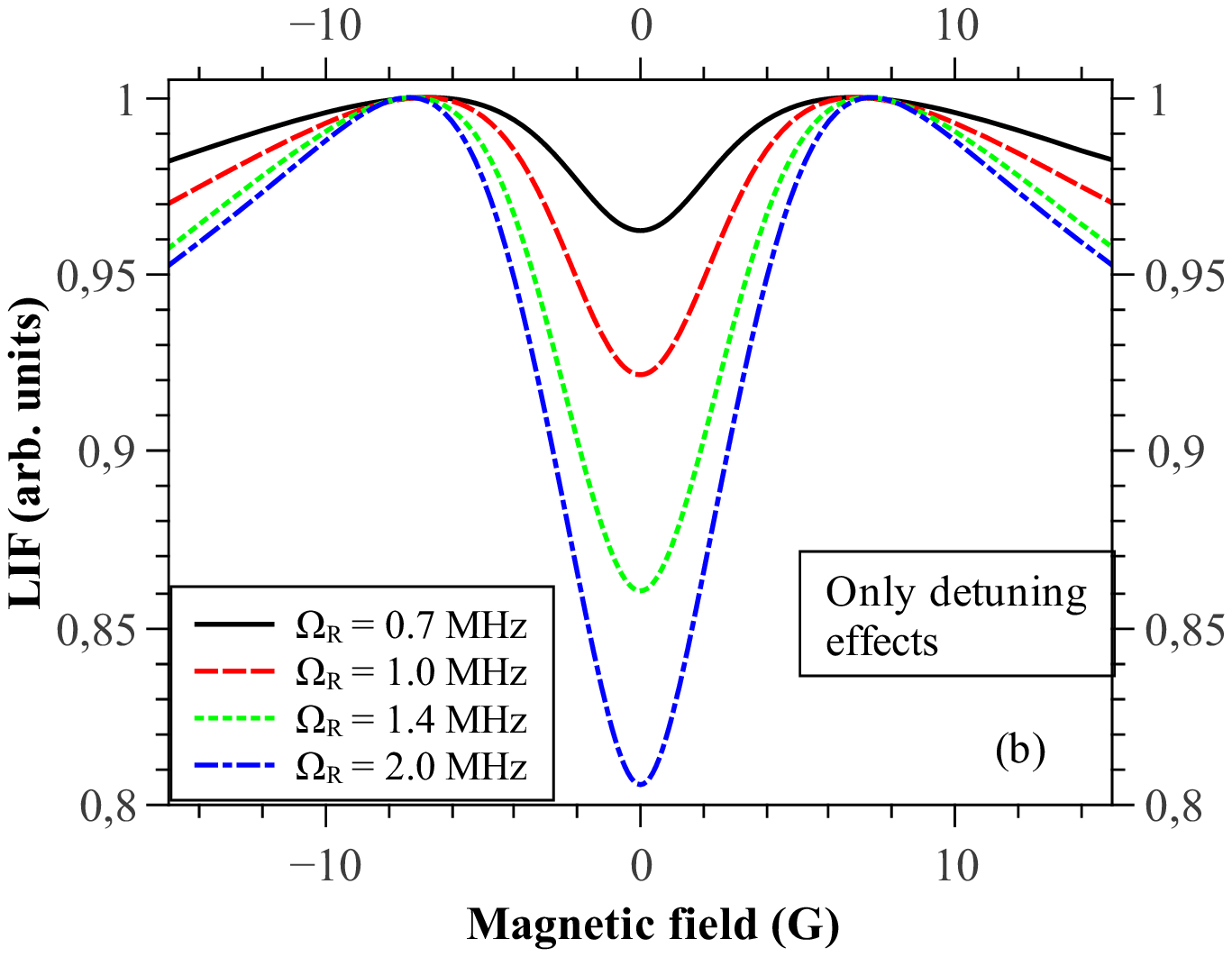}}\\%
\resizebox{.5\textwidth}{!}{\includegraphics[width=.5\textwidth]{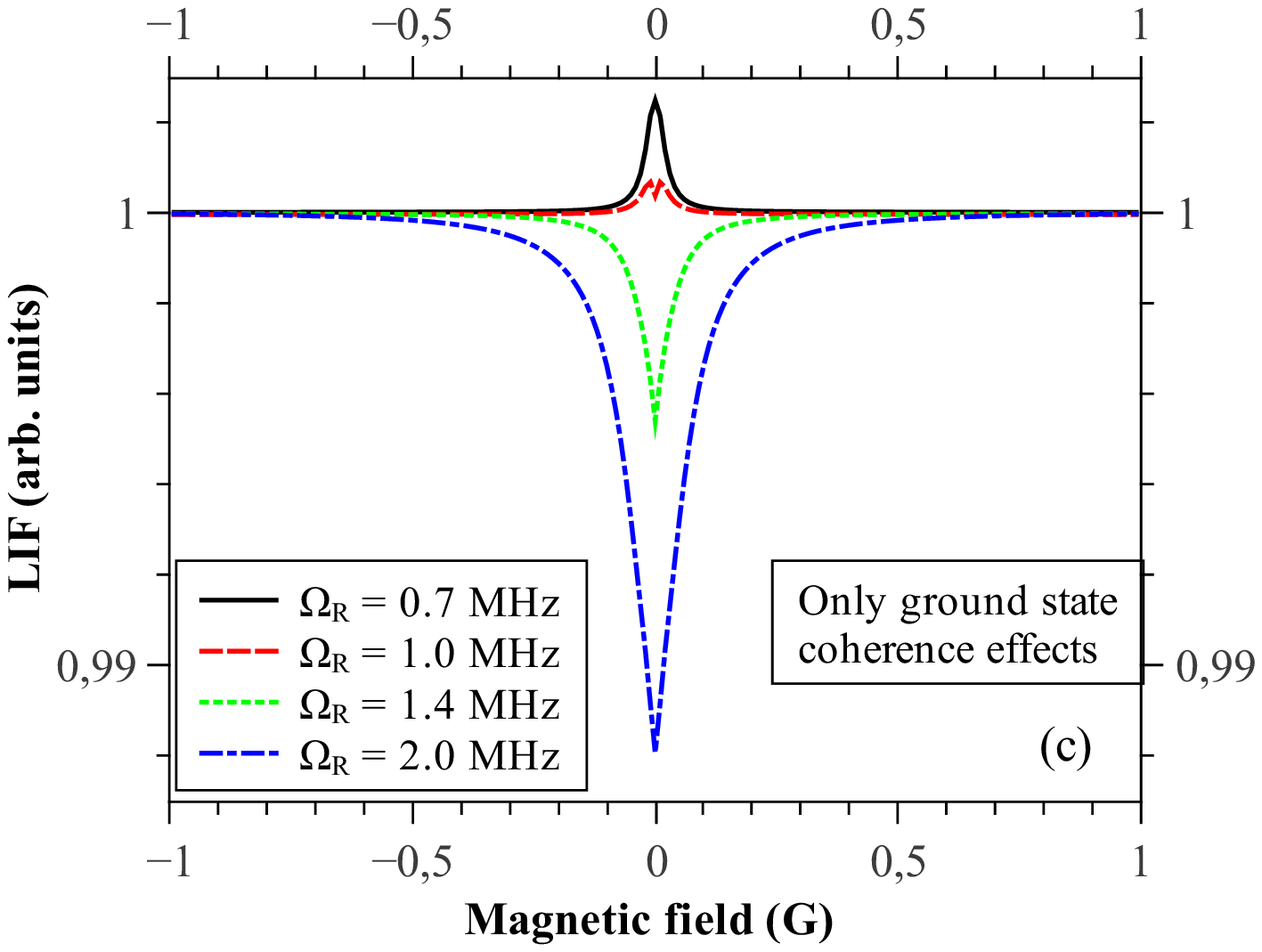}}%
\resizebox{.5\textwidth}{!}{\includegraphics[width=.5\textwidth]{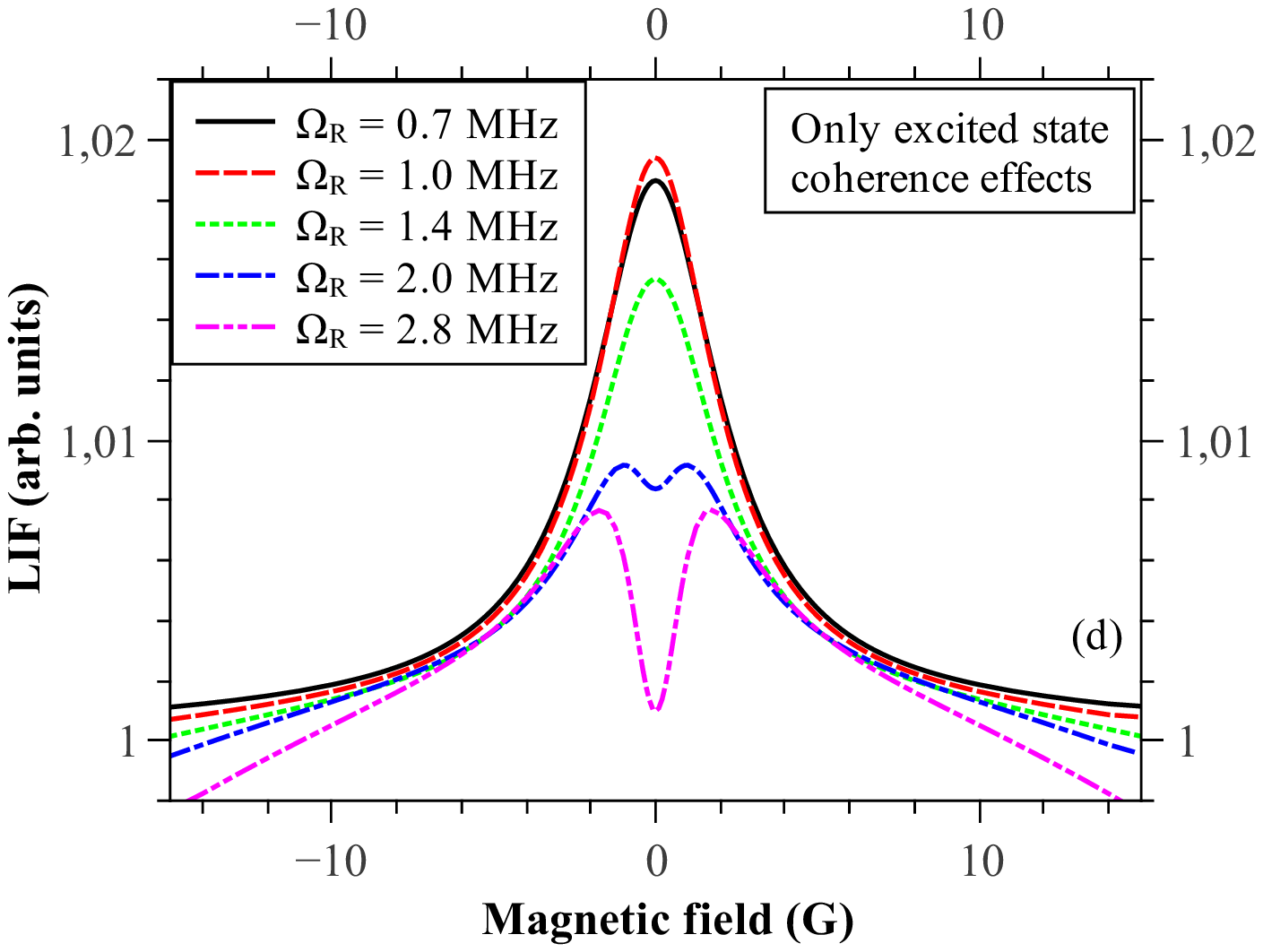}}%
\caption{\label{fig:effects} (colour online) Theoretical calculations of LIF versus magnetic field $B$ for the $F_g=2\longrightarrow F_e=3$ transition of $^{87}$Rb with different physical effects taken into account: (a) all effects taken into account,  (b) detuning effects only, (c) ground state coherence effects only, (d) excited state coherent effects only.  Note the different scales! The parameters used in the simulation were as follows: $\gamma=0.019$ MHz, $\Delta \omega_{Laser}= 2$ MHz, $\sigma_{Doppler}=216$ MHz, $D_{Step} \approx 1.73$ MHz}
\end{figure*}

In order to study how different physical effects influence those features of the signal that appear at different scales of the magnetic field, we used the same theoretical model, while turning different physical processes ``on" and ``off". Three processes were considered: destruction of ground-state coherences by the magnetic field, destruction of excited-state coherences by the magnetic field, and the ``Zeeman magnetic scanning effect", which involved optical transitions between different Zeeman sublevels that come into resonance with the laser radiation as a function of the magnetic field strength and the 
atomic velocity. The results are 
shown in Fig.~\ref{fig:effects}. When all effects were included, we obtained structures on the scale of 100 mG, several Gauss, and several tens of Gauss [see Fig. \ref{fig:effects}~(a)]. The latter two are, as we will show later, caused by detuning effects as the hyperfine levels are split in the external magnetic field, and we will refer to these features as the ``wide structure''.

When the effect of the changing magnetic field on the coherences was neglected, which was done by setting the third term in (\ref{eq:zc}) to zero for both ground and excited states [Fig. \ref{fig:effects}~(b)], the small, narrow peaks disappeared completely, whereas the other structures remained largely, but not completely, unchanged.
In order to consider only the ground-state coherence effects, the excited-state coherences were decoupled from the magnetic field by setting the third term of (\ref{eq:zcee}) to zero, and the 
detuning effects were "turned off" by taking the term $\omega_{g_ie_j}$ in the denominator of (\ref{eq:xi}) to be independent of the magnetic field and keeping its value at the value it has at $B=0$. Only the narrow structure was reproduced when only the magnetic field's destruction of the ground-state coherence effects were taken into account in Fig. \ref{fig:effects}~(c). The results shown in Figures \ref{fig:effects}~(b) and \ref{fig:effects}~(c) clearly attribute the narrow structure to the ground state coherences and their destruction by the magnetic field. The flip-over of the narrow structure that can be seen in Figs.~\ref{fig:profiles}~(a)--(b) and \ref{fig:effects}~(a) and (c) while increasing the laser power density has been explained earlier~\cite{Auzinsh:2012}. At the same time the resonance with a width of several Gauss in Fig. \ref{fig:effects}~(b) is seen to be related to detuning effects, which where the only ones considered in that calculation.

When only the excited-state coherent effects were taken into account in a similar way, a structure with negative second derivative and a width of several Gauss appeared; the contrast was only one or two percent [Fig. \ref{fig:effects}~(d)]. The structure had the same characteristic width ($\Gamma \approx \omega_{\Delta m=2}$) as the linear Hanle effect of the excited state~\cite{Strumia}. 
The linear Hanle effect cannot be observed in our experiment as it requires discrimination of the polarization components of the LIF, and so we attribute this structure to the nonlinear Hanle effect of the excited state. Calculations at several Rabi frequencies showed that the peak associated with this effect became smaller as the Rabi frequency changed from $1.0$~MHz to $2.0$~MHz. Moreover, at $2.0$~MHz another small dip with positive second derivative appeared inside the peak at zero magnetic field; a further increase in Rabi frequency indicates a similar behaviour, though on a different scale as in Fig.~\ref{fig:effects}~(c) (the effects produced by the destruction of ground state coherences). In any case, the calculations show that excited state coherences play no role in the narrow structure.

\begin{figure*} 
\resizebox{\textwidth}{!}{\includegraphics[width=\textwidth]{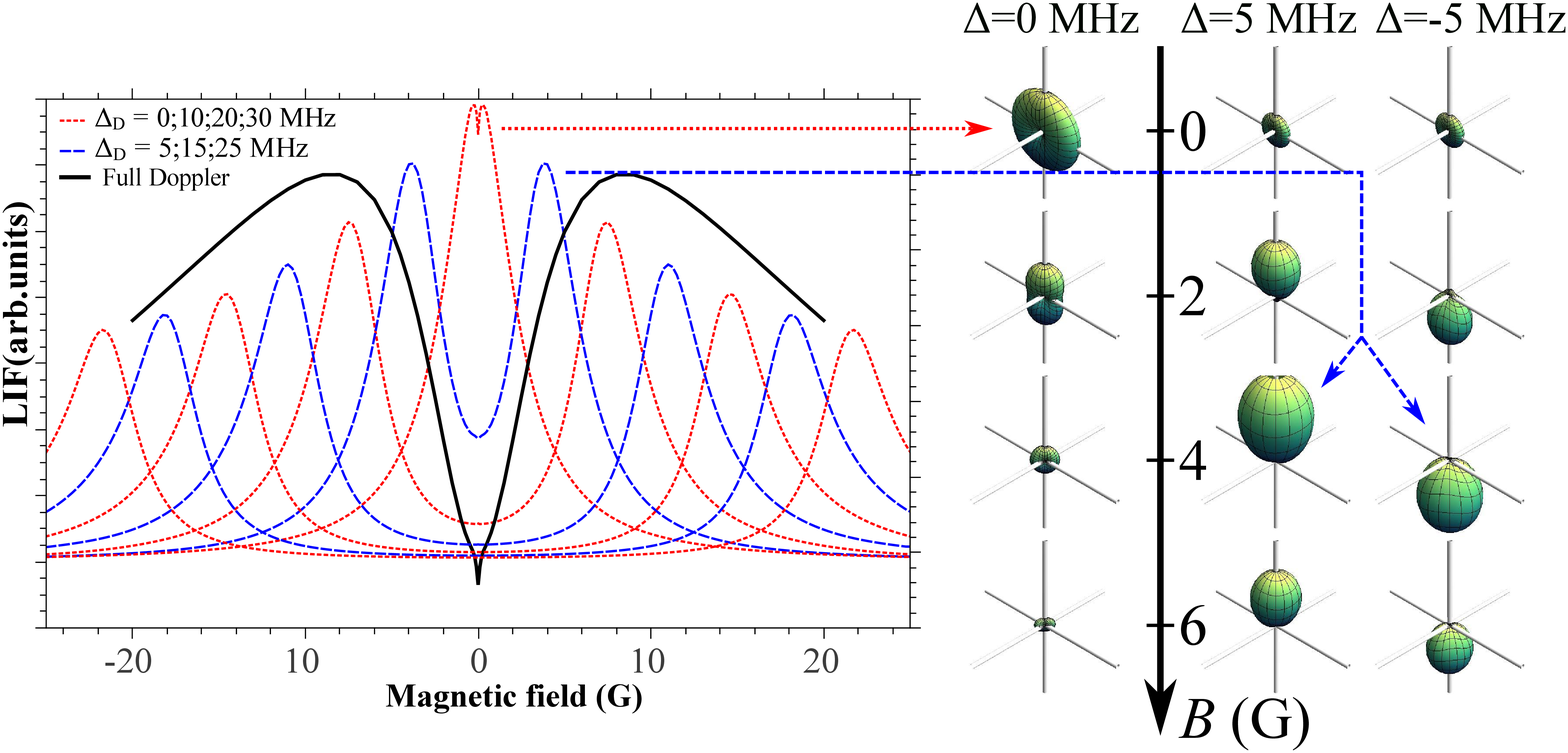}}%
\caption{\label{fig:vsop} Decomposition of a magneto-optical signal into a superposition of signals from different velocity groups and at different magnetic fields. Left panel: The solid, black line shows the magneto-optical signal as it would be observed in a vapor cell at room temperature. The dashed and dotted lines show the signals for the different velocity groups that make up the room temperature velocity distribution. Right panel: Distribution of the atomic angular momentum at different values of the magnetic field $B$ for the velocity groups in resonance at a (Doppler) detuning of 0 MHz, 5 MHz, and -5MHz.  }
\end{figure*}

The main origin of the wide structure can be understood by considering Fig.~\ref{fig:vsop}. The left panel shows how a magneto-optical signal can be decomposed into contributions from different velocity groups. The solid black line represents the signal of a vapour at room temperature and is formed from an average over all the velocity groups in the Doppler profile. The dashed and dotted lines represent contributions from different velocity groups. One can see that the superposition of the contributions from the dashed and dotted lines would yield a shape similar to the black line. The right panel explains why each velocity group has its own shape. The laser is assumed to be on resonance at zero magnetic field with a group of atoms that is stationary with respect to the propagation direction of the laser radiation ($v_x = 0$) for the $F_g = 2 \rightarrow F_e = 3$ transition. All other velocity groups therefore interact negligibly with a laser field that is detuned by the Doppler shift. As the magnetic field is applied, all 
magnetic sublevels shown in Fig. \ref{fig:probabilities}, except those with $m=0$, are shifted as a result of the Zeeman effect. We may say that a magnetic scanning is performed by bringing into resonance a group of atoms with some velocity $v_x = v(B)$. The function $v(B)$ in general is nonlinear and is explicitly determined by the nature of the (nonlinear) Zeeman effect. As a result of the magnetic scanning, the shapes of the angular momentum distributions induced by the laser radiation differ as a function of magnetic field for each velocity group, which can be explicitly shown by the angular momentum probability surfaces~\cite{Auzinsh:1997,Rochester:2001} for the excited state. When the angular momentum probability surfaces are drawn, only the $F_e=3$ hyperfine level is taken into account, as other hyperfine levels are far away from resonance for the magnetic field values and velocity groups shown in Fig. \ref{fig:vsop}, and their input populations are negligible. We may anticipate from Fig. \ref{fig:vsop} and the preceding discussion that, at a particular magnetic field value, some group of atoms with corresponding velocities becomes effectively oriented in either the positive or the negative direction of the axis along which 
the magnetic field is applied. Further, the whole ensemble of atoms becomes aligned along the same axis at magnetic values that produce the LIF maxima around $\pm 10$ Gauss.

\section{Conclusion}
Nonlinear magneto-optical resonances from the $D_2$ line of $^{87}$Rb have been studied experimentally and theoretically up to magnetic field values of 40~G. The theoretical model was based on the optical Bloch equations and included the coherence properties of the laser radiation, all adjacent hyperfine transitions, the mixing of magnetic sublevels in the external magnetic field, and the Doppler effect. The model described the experimentally measured signals very well. By removing individual physical processes from the model, it was possible to deduce the physical origin of the different features observed in the signals. As expected, the narrow structure was related to coherences among ground-state Zeeman sublevels induced by the exciting laser radiation. Coherences among excited-state sublevels were found to have a small effect on signals at magnetic field scales of several Gauss. The origin of the wide structure was explained in terms of contributions from different 
velocity groups. With these results, it is possible to understand the origin of the variation in LIF as a function of magnetic fields in the range 
up to at least several tens of Gauss.  

We may conclude that the results of this study emphasize the necessity to incorporate a number of processes in a theoretical model that aims to provide a quantitative description of magneto-optical effects. The most important of these effects are 1) the Doppler effect, 2) the magnetic scanning, and 3) the change in the transition probabilities due to the magnetic mixing of the hyperfine levels, which can reach $30\%$ for $^{87}$Rb $D_2$ excitation at $B=40$~G. Although each of the processes can be treated separately to obtain an analytical description, in order to have an accurate description that is valid over a wider range of laser power densities and magnetic field values, one has to treat all the processes simultaneously. On the other hand, a numerical model that incorporates a number of processes can be used to estimate limiting conditions for various approximations used in analytical models in the way described above.

\ack
The contributions of Artis Kruzins to the experiments is highly appreciated. We are grateful to the Latvian State Research Programme No. 2010/10-
4/VPP-2/1 and the NATO Science for Peace project CBP.MD.SFPP.983932, ``Novel Magnetic Sensors and Techniques for Security Applications'' for financial support.

\section*{References}

\bibliographystyle{unsrt}
\bibliography{rubidium}

\end{document}